\documentstyle[12pt]{article}        
\newlength{\dinwidth}                       
\newlength{\dinmargin}                      
\begin{document}
\vspace*{1cm}
\begin{center}  \begin{Large} \begin{bf}
Binding effects and nuclear shadowing \\ 
  \end{bf}  \end{Large}
D.\ Indumathi$^a$ and W.\ Zhu$^b$ \\
  \vspace*{5mm}
  \begin{large}
  \end{large}
\end{center}
$^a$ Institut f\"ur Physik, Universit\"at Dortmund,
D-44221, Germany 

\noindent $^b$ Department of Physics, East China Normal University,
Shanghai 200 062, P.R.China 

\begin{quotation}
\noindent
{\bf Abstract:}
The effects of nuclear binding on nuclear structure functions have so
far been studied mainly at fixed target experiments, and there is
currently much interest in obtaining a clearer understanding of this
phenomenon. We use an existing dynamical model of nuclear structure
functions, that gives good agreement with current data, to study this
effect in a kinematical regime (low $x$, high $Q^2$) that can possibly
be probed by an upgrade of {\sc hera} at {\sc desy} into a nuclear
accelerator. 
\end{quotation}


\noindent The ratio of the structure functions of bound and free
nucleons is smaller than one at $x<0.1$; this has been observed
previously and is called nuclear shadowing \cite{EMC1}. Nuclear
shadowing and its scaling proporties are generally regarded as the
shadowing effect arising from gluon recombinations in the partonic
model. A surprising fact of the {\sc hera} data is the rapid rise
of the structure function $F_2$ of the proton as $x$ decreases.
The expected shadowing effect of gluon recombinations is not
visible at least down to $x \sim 10^{-3}$ \cite{hera}. On the other
hand, one of us (WZ) \cite{wz1} has pointed out that the effect of
shadowing due to gluon recombinations on a steep gluon distribution
will be weakened by momentum conservation; in particular, gluon fusion
can be neglected in the QCD nonlinear evolution equation in the small-$x$
region where the gluon density rises like the Lipatov $x^{-1/2}$
behavior. Obviously, reconsideration of the partonic shadowing model is
necessary. We have thus evolved a new approach to nuclear shadowing,
which explains available data without needing Glauber rescattering
\cite{IZ}.  On the other hand, there is a strong likelihood of {\sc hera}
being upgraded to a nuclear accelerating machine \cite{Workshop}. We
therefore apply our model and obtain predictions for the nuclear
structure functions in the kinematical regime of the {\sc hera} machine. 

\paragraph{The Model}:
We quickly review the model. We consider the DIS process in the Breit
frame, where the exchanged virtual boson is point-like and the target
consists of partons. The $z$-component of the momentum of the struck
quark is flipped in the
interaction. Hence, due to the uncertainty principle, a struck quark
carrying a fraction $x$ of the nucleon's momentum, $P_N$, during the
interaction time $\tau_{\rm int}=1/\nu$, will be off-shell and
localized longitudinally to within a potentially large distance
$\Delta z\sim 1/(2xP_N)$, which may exceed the average two-nucleon
separation $D_A$ for a small enough $x<x_0=1/m_ND_A$.

The struck sea quark with its parent will return to its initial
position within $\tau_{\rm int}$ if the target is a free nucleon. However,
in a bound nucleon target, it can interact with other nucleons in the
nucleus and so loses its energy-momentum. Since it can be randomly
distributed outside the target nucleon, and interacts incoherently with
the rest of the nucleus, we regard this effect as an additive (second)
binding effect rather than as a Glauber rescattering.

A simple way of estimating the second binding effect is to connect this
new effect with the traditional binding effect, which influences the
parton input distributions at the starting point, $Q^2 = \mu^2$, of
the QCD evolution. At such low scales, we picture the nucleon as being
composed of valence quarks, gluons, and mesonic sea quarks. For
example, we identify the GRV (LO) parametrisations \cite{GRV} as the
input parton distributions of the free nucleon at $\mu^2=0.23$\ GeV${}^2$.

We consider that the attractive potential describing the nuclear force
arises from the exchange of scalar mesons. Hence the energy required for
binding is taken away solely from the mesonic component of the
nucleon, and not from its other components. We identify this with the
sea quarks (and antiquarks) in the nucleon. Therefore, we assume that
the nuclear binding effect only reduces the sea distributions of the
nucleon at $Q^2=\mu^2$. 

For a binding energy, $b$, per nucleon, this corresponds to the
reduction of the bound nucleon sea densities from the free-nucleon
value, $S_N(x, \mu^2)$, given by GRV at $Q^2=\mu^2$ to
$$
\begin{array}{rcl}
S_A (x, \mu^2) & = & K(A)S_N(x,\mu^2) \\
 & = & \left(1-{\displaystyle 2b \over \displaystyle M_N\langle
        S_N(\mu^2)\rangle_2}\right) S_N(x,\mu^2)~.
\end{array}
\eqno(1)
$$
Here $\langle S_N\rangle_2$ is the momentum fraction (second moment)
of the sea quarks and we assume that the decrease in number of sea
quarks due to the binding effect is proportional to their density.

We assume that the energy loss of sea quarks, $U_s(Q^2)$, due to the
second binding effect is also proportional to the density,
$$
U_s(Q^2) = \beta M_N\int_{0}^{x_0}xS_A(x,Q^2)
    \simeq\beta M_N\langle S_A(Q^2)\rangle_2~,
\eqno(2)
$$
and the strength of this interaction is similar to eq. (1), viz.,
$$
\beta=\frac{U_s(Q^2)}{M_N\langle S_A(Q^2)\rangle_2}
       =\frac{U(\mu^2)}{M_N\langle S_N(\mu^2)\rangle_2}~,
\eqno(3)
$$
$U(\mu^2)=a_{\rm vol}/6$ being the binding energy between each pair
of nucleons. In consequence, we have, due to the second binding effect,
a depletion of the sea quarks, given by
$$
S_A(x,Q^2)-S_A'(x,Q^2)=\beta S_A(x,Q^2)~.
\eqno(4)
$$
Combining the above mentioned two kinds of binding effects and the
swelling effect on a bound nucleon, which was discussed in
ref.\ \cite{wz2}, we are able to explain recent data on the EMC effect
in a broad kinematical region using only a few fundamental nuclear
parameters \cite{IZ}.

\paragraph{Predictions for HERA}:
Our model predicts correctly the $A$-, $Q^2$ and $x$-dependences of the
nuclear structure functions (and not only the ratio of cross sections)
measured by the NMC in the ranges of $x < 0.8$, $Q^2 > 0.5$ GeV${}^2$.
However, being a fixed target experiment, the available $Q^2$ range is
limited. It would therefore be interesting, as well as instructive,
to obtain more results and predictions in the small-$x$ region, and in
a larger $Q^2$ range.

\begin{figure}[htb]
\vskip 7truecm

\includegraphics{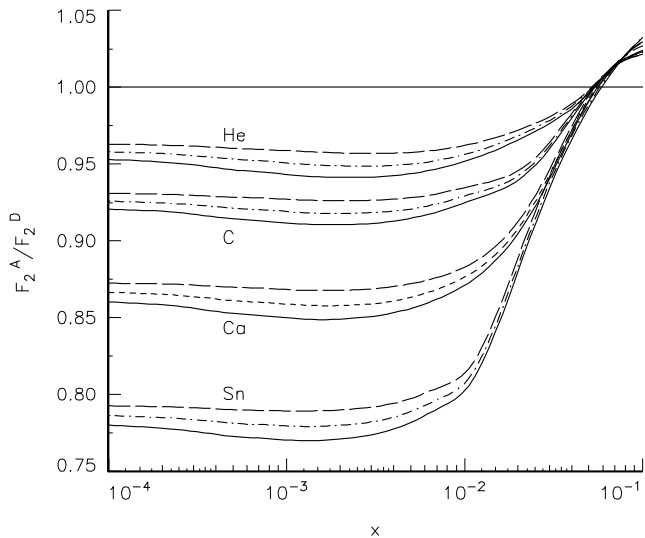}
\vspace{0.3cm}

\caption[dummy]{\it 
The structure function ratios as functions
of $x$ and $Q^2$ for He/D, C/D, Ca/D, and Sn/D. The full,
broken, and long-dashed curves correspond to $Q^2 = 4, 30$, and
$100$~${\rm GeV}^2$ respectively.} 
\label{fig1}
\end{figure}

We have done this for the case of a few typical nuclei, the results for
which are shown in Fig.\ \ref{fig1}. We see that the small-$x$ shadowing
is weakly dependent on $Q^2$ and is dominated by the second binding
effect. Due to this, there is hardly any $x$-dependence, as saturation
has already set in by $x \sim 10^{-2}$. However, the magnitude of the
shadowing is large for heavier nuclei. 

We therefore conclude that it would be definitely worthwhile to make the
effort to study such processes at {\sc hera} in an attempt to understand
better the nature of the binding of nucleons in nuclei. 

\vspace{0.5cm}

\end{document}